\documentclass[aps,pra,reprint,amsmath,amssymb,superscriptaddress]{revtex4-2}

\usepackage{graphicx}
\usepackage{url}
\usepackage{hyperref}

\begin{document}
	
	\title{Insightful Approach to Quantum Noise Suppression Below the Standard Quantum Limit Using a Single Mirror and Beam Splitter}
	
	\author{Sun-Hyun Youn}
	\email{sunyoun@jnu.ac.kr}
	\affiliation{Department of Physics, Chonnam National University, Gwangju 61186, Korea}
	
	\date{\today}
	
	\begin{abstract}
		When a coherent electromagnetic wave passes through a beam splitter (BS), it is equally divided into two parts. However, the quantum noise associated with the resulting coherent states, while their amplitudes are reduced by half, remains fundamentally constrained by the quantum noise limit and is independent of intensity. By placing a mirror at the unused input port of the BS, a standing wave is formed near the mirror, which influences the vacuum fluctuations of the coherent state at the BS output. Using both semi-classical and quantum mechanical approaches, we calculate the vacuum fluctuations induced by the mirror and show that the vacuum noise originating from the mirror periodically vanishes at the BS output. Leveraging this effect, we demonstrate that the vacuum fluctuations of the light split by the BS can be reduced below the standard quantum limit. Furthermore, by incorporating feedback mechanisms, the vacuum fluctuations of the electromagnetic field at the other output port can also be suppressed below the quantum noise limit at certain locations. 
		Remarkably, our method achieves this noise suppression using only a single mirror and a beam splitter, without requiring nonlinear optical processes or complex experimental apparatus. The key insight underlying this simplicity is the recognition that quantum noise can be controlled by engineering the spatial ``mode'' of light---a concept fundamentally distinct from manipulating its quantum ``wave function.'' 
		These findings offer key insights into the control of electromagnetic noise, with broad implications for experiments involving quantum noise suppression.
	\end{abstract}
	
	\maketitle

	\section{Introduction}
	
	The reduction of noise in electromagnetic waves is a critical challenge in both fundamental quantum physics and applied technologies, such as precision measurement and quantum information processing. Noise in electromagnetic waves can be broadly categorized into classical and quantum contributions. Classical noise, arising from environmental fluctuations or technical imperfections, can often be mitigated through feedback mechanisms. In such systems, a portion of the emitted electromagnetic wave is measured, and any detected fluctuations are actively corrected via feedback loops \cite{Zhang2017}. For instance, in semiconductor lasers, feedback techniques have been employed to reduce amplitude noise by extracting a portion of the light and applying corrective measures \cite{Yamamoto1986}. However, these methods are fundamentally limited by the quantum noise floor, often referred to as the standard quantum limit (SQL). This limit arises because each measurement process introduces additional vacuum fluctuations, preventing noise reduction below the SQL \cite{Youn1994,Zhang2017}. As a result, feedback-based approaches, while effective for classical noise, are insufficient for addressing quantum noise, necessitating alternative strategies.
	
	To overcome the quantum noise limit, significant research has focused on the generation of non-classical states of light, particularly squeezed states, which exhibit reduced noise in one quadrature of the electromagnetic field at the expense of increased noise in the conjugate quadrature \cite{Slusher1985}. The concept of squeezed states, first experimentally demonstrated in the 1980s by Slusher \textit{et al.} using four-wave mixing near an atomic resonance \cite{Slusher1985}, has since become a cornerstone of quantum optics. Squeezed states offer the potential to reduce noise below the SQL, making them invaluable for applications requiring ultra-low noise electromagnetic waves. Over the past decades, numerous techniques for generating squeezed states have been developed, including optical parametric amplification \cite{Wu1986}, degenerate four-wave mixing \cite{Shelby1986}, and the use of nonlinear optical crystals \cite{Wu1986}. These methods have been applied in various contexts, such as improving the sensitivity of interferometric measurements \cite{Caves1981}, enhancing quantum communication protocols \cite{Braunstein2005}, and enabling quantum metrology \cite{Giovannetti2004}. More recent advances have explored the generation of squeezed states in integrated photonic systems \cite{Cui2021}, cavity optomechanical systems \cite{Lee2020}, and superconducting circuits \cite{Castellanos2008}, highlighting the versatility and ongoing relevance of this field.
	
	A crucial conceptual distinction, often overlooked in the literature, lies at the heart of understanding and controlling quantum noise: the difference between the quantum ``wave function'' of light and the spatial ``mode'' it occupies. The wave function describes the quantum state of the electromagnetic field, while the mode defines the spatial structure governed by Maxwell's equations and boundary conditions. Quantum noise is fundamentally determined by the field operators ($\hat{a}$, $\hat{a}^\dagger$) acting on specific spatial modes, not merely by the quantum state itself. It is well established that a mode's spatial structure directly shapes the vacuum fluctuations within it. For example, the standing-wave mode formed between two mirrors imposes a corresponding standing-wave structure on the vacuum fluctuations, creating nodes where fluctuations are exceptionally suppressed---an effect experimentally verified since the 1980s in studies of the Casimir effect and cavity quantum electrodynamics \cite{Casimir1948,Milton2001}. This insight suggests that by engineering the spatial mode structure through simple boundary conditions, one can achieve quantum noise suppression without resorting to nonlinear optical operations on the wave function.
	
	The need for low-noise electromagnetic waves is particularly evident in precision measurement experiments, where quantum noise imposes fundamental limits on sensitivity. A prominent example is the Laser Interferometer Gravitational-Wave Observatory (LIGO), which relies on ultra-precise interferometry to detect gravitational waves \cite{Aasi2015,Tse2019}. In such experiments, quantum noise, including shot noise and radiation pressure noise, limits the achievable signal-to-noise ratio, making the use of squeezed states essential for enhancing detection capabilities \cite{Schnabel2010}. Beyond gravitational wave detection, low-noise electromagnetic waves are crucial for applications in quantum computing, where noise reduction is necessary to maintain coherence \cite{Lu2023}, and in optical communication, where squeezed states can improve channel capacity \cite{Derkach2020}.
	
	In this work, we propose a novel approach to reduce electromagnetic wave noise below the vacuum fluctuation level by utilizing a mirror-based beam splitter to generate squeezed states. Our method leverages the interference properties of beam splitters to manipulate quantum fluctuations, offering a practical and scalable solution for noise reduction.
	
	It is well established that any network of beam splitters, phase shifters, and mirrors acting on vacuum plus coherent inputs remains a mixture of coherent and vacuum modes~\cite{Caves1982}. However, the key contribution of this paper lies in modifying the spatial characteristics of the vacuum modes by imposing specific boundary conditions. In quantum optics, the quantum mechanical state of light is described by quantum mechanical operators acting on modes determined by the classical Maxwell equations~\cite{Fabre}. Our objective is not to alter the framework of quantum mechanical operators but to modify the properties of the modes on which these operators act.
	
	The interaction of electromagnetic fields at a beam splitter is a fundamental concept in quantum optics, crucial for phenomena ranging from interferometry to quantum information processing. This work explores a specific scenario where a coherent traveling wave, $\hat{E}_b$, mixes with a standing wave in its vacuum state, $\hat{E}_c$, at a beam splitter. The resulting fields, $\hat{E}_1$ and $\hat{E}_2$, inherit characteristics from both inputs, leading to unique spatial and temporal noise profiles. We aim to rigorously characterize these output fields using both semi-classical and fully quantum frameworks, propose a practical measurement technique, and demonstrate the possibility of noise reduction below the standard quantum limit.
	
	This paper is organized as follows: First, we review the theoretical framework of quantum noise and squeezed states. Next, we describe the experimental implementation of our mirror-based beam splitter system. Finally, we present our results and discuss their implications for future applications in quantum technologies.

	\section{Semi-classical Description of Fields and Noise}
	
	\subsection{Semiclassical Model of Coherent States}
	
	The semiclassical model of a coherent state with a specific frequency consists of a spatially well-defined propagating mode $E_0 \sin(\omega t - kz + \theta)$ combined with vacuum fluctuations. At a given spatial point, the vacuum fluctuations can be modeled as a superposition of forward- and backward-propagating waves with the same frequency as the coherent state:
	\begin{multline}
		E_{\text{vac}} = \sum_{\phi_{b1}} b_F(\phi_{b1}) \sin(\omega t - kz + \phi_{b1}) \\
		+ \sum_{\phi_{b2}} b_B(\phi_{b2}) \sin(\omega t + kz + \phi_{b2}),
	\end{multline}
	where $b_F$ and $b_B$ satisfy the ensemble average conditions $\langle b_F \rangle = \langle b_B \rangle = 0$, and $\langle b_F^2 \rangle = \langle b_B^2 \rangle$ characterizes the magnitude of vacuum fluctuations.
	
	The ensemble-averaged electric field at a specific time and position is:
	\begin{align}
		\langle E \rangle &= \frac{1}{(2\pi)^2} \int_0^{2\pi} \int_0^{2\pi} \bigg[E_0 \sin(\omega t - kz + \theta) \nonumber \\
		&\quad + b_F(\phi_{b1}) \sin(\omega t - kz + \phi_{b1}) \nonumber \\
		&\quad + b_B(\phi_{b2}) \sin(\omega t + kz + \phi_{b2})\bigg] \nonumber \\
		&\quad \times d\phi_{b1} d\phi_{b2} \nonumber \\
		&= E_0 \sin(\omega t - kz + \theta).
	\end{align}
	
	The mean squared electric field is:
	\begin{equation}
		\langle E^2 \rangle = \frac{1}{2}\left(\langle b_B^2 \rangle + \langle b_F^2 \rangle + E_0^2 \sin^2(\omega t - kz + \theta)\right).
	\end{equation}
	
	Therefore, the field variance is:
	\begin{equation}
		(\Delta E)^2 = \langle E^2 \rangle - \langle E \rangle^2 = \frac{1}{2}\left(\langle b_F^2 \rangle + \langle b_B^2 \rangle\right),
	\end{equation}
	which is independent of the coherent amplitude $E_0$.
	
	\subsection{Beam Splitter Configuration with Mirrors}
	
\begin{figure}[htb]
		\centering
		\includegraphics[width=0.6\linewidth]{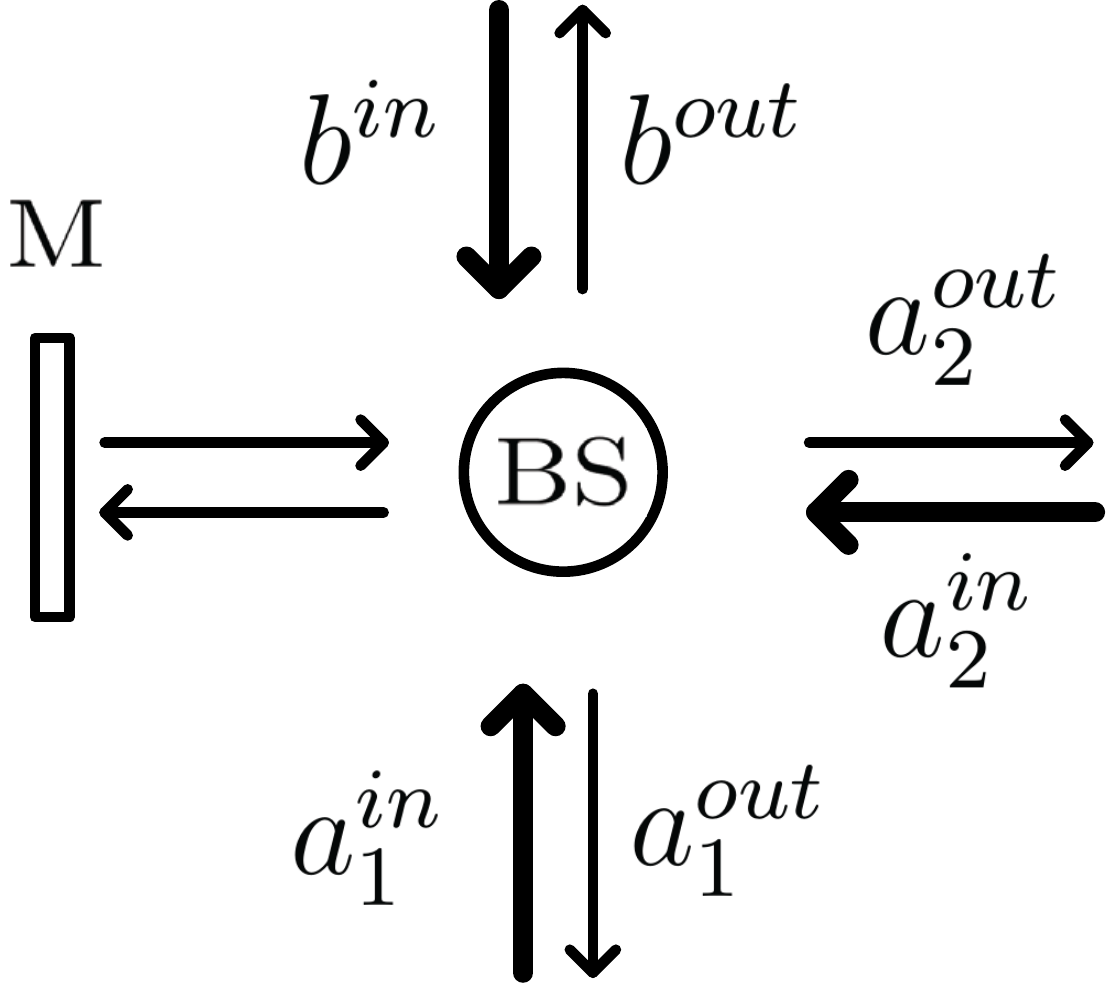}
		\caption{Schematic of the beam splitter (BS) setup. One input port of the beam splitter is the $b$ mode, and a mirror is placed at the other port. The modes $a_1$ and $a_2$ denote the two output ports of the BS.}
		\label{bsSetCLA}
	\end{figure}

	Consider the beam splitter configuration shown in Fig.~\ref{bsSetCLA}. The beam splitter relations for the field arriving from the mirror (denoted as $c$) are:
	\begin{align}
		a_1^{\text{out}} &= \sqrt{T} b^{\text{in}} + \sqrt{R} c, \\
		a_2^{\text{out}} &= \sqrt{R} a_1^{\text{in}} + \sqrt{T} a_2^{\text{in}}.
	\end{align}
	
	Since the electromagnetic wave from the mirror to the beam splitter is determined by $a_1^{\text{in}}$ and $a_2^{\text{in}}$, the output fields satisfy:
	\begin{align}
		b^{\text{out}} &= \sqrt{T} a_1^{\text{in}} - \sqrt{R} a_2^{\text{in}}, \\
		a_1^{\text{out}} &= \sqrt{T} b^{\text{in}} - R a_1^{\text{in}} - \sqrt{RT} a_2^{\text{in}}, \\
		a_2^{\text{out}} &= -\sqrt{R} b^{\text{in}} - T a_2^{\text{in}} - \sqrt{RT} a_1^{\text{in}}.
	\end{align}
	
	When a coherent state enters port $b$, the electric field at port $a_1$ is:
	\begin{align}
		E_1 &= \sqrt{T} \Big( E_0 \sin[\omega t - k(z_1 + L_0)] \nonumber \\
		&\quad + \sum_{\phi_{b}} b_F(\phi_b) \sin[\omega t - k(z_1 + L_0) + \phi_b] \Big) \nonumber \\
		&\quad +\sum_{\phi_{1}} \big\{ a_{1n} (\phi_1) \sin[\omega t + kz_1 + \phi_1] \nonumber \\
		&\quad - R a_{1n}(\phi_1) \sin[(\omega t - kz_1) + \phi_1] \big\} \nonumber \\
		&\quad - \sum_{\phi_{2}}\sqrt{RT} a_{2n}(\phi_2 )  \sin[(\omega t - kz_1) + \phi_2],
	\end{align}
	where $z_1$ is the optical path length from the mirror, $L_0$ is related to the optical path length of the coherent state, and $a_{1n}$, $a_{2n}$ model the vacuum modes propagating from ports $a_1$, $a_2$ to the beam splitter, with $\langle a_{1n} \rangle = \langle a_{2n} \rangle = 0$ and $\langle a_{1n}^2 \rangle = \langle a_{2n}^2 \rangle$ characterizing the vacuum noise magnitude.
	
	The ensemble average of this field is:
	\begin{equation}
		\langle E_1 \rangle = E_0 \sqrt{T} \sin[\omega t - k(L_0 + z_1)].
	\end{equation}
	
	The mean squared field is:
	\begin{align}
		\langle E_1^2 \rangle &= \frac{1}{2} \bigg[-2a_{1n}^2 R \cos(2kz_1) + a_{1n}^2(R^2 + 1) \nonumber \\
		&\quad + T(a_{2n}^2 R + E_0^2 + b_F^2) \nonumber \\
		&\quad - E_0^2 T \cos(2\omega t - 2k(L_0 + z_1))\bigg].
	\end{align}
	
	The field variance at output port $a_1$ is:
	\begin{align}
		\langle \Delta E_1^2 \rangle &= \frac{1}{2} \Big[ T(a_{2n}^2 R + b_F^2) \nonumber \\
		&\quad -2a_{1n}^2 R \cos(2kz_1) + a_{1n}^2(R^2 + 1) \Big].
	\end{align}
	
	Similarly, the field variance at output port $a_2$ is:
	\begin{align}
		\langle \Delta E_2^2 \rangle &= \frac{1}{2} \Big[R(a_{1n}^2 T + b_F^2) \nonumber \\
		&\quad - 2a_{2n}^2 T \cos(2kz_2) + a_{2n}^2(T^2 + 1)\Big].
	\end{align}
	
	\subsection{Conservation of Field Fluctuations}
	
	An important observation is that the initial coherent state field variance was $\frac{1}{2}(b_F^2 + b_B^2)$, where $b_B$ is associated with the vacuum fluctuations in the mode related to $b^{\text{out}}$. Using the relation $b^{\text{out}} = \sqrt{T}a_1^{\text{in}} - \sqrt{R}a_2^{\text{in}}$, we find:
	\begin{equation}
		(\Delta E)^2 = \frac{1}{2}(b_F^2 + a_{2n}^2 R + a_{1n}^2 T). \label{delCo}
	\end{equation}
	
	The field variances can be rewritten as:
	\begin{align}
		\langle \Delta E_1^2 \rangle &= T (\Delta E)^2 + 2a_{1n}^2 R \sin^2(kz_1), \\
		\langle \Delta E_2^2 \rangle &= R (\Delta E)^2 + 2a_{2n}^2 T \sin^2(kz_2).
	\end{align}
	
	This demonstrates that when a lossless beam splitter divides a coherent state into two outputs, both the field amplitude and the coherent state noise are divided accordingly. The vacuum noise entering from the unused port ensures that the total field fluctuation maintains a constant value, confirming this behavior even in the presence of mirrors in the beam splitter configuration.
	
	\subsection{Predicted Noise Characteristics of $E_1(z_1)$}
	
	The noise in $E_1$ is determined by contributions from both input ports of the beam splitter. At the port where $E_0$ enters, the incoming noise is represented by the vacuum fluctuation $b_F$, which has a uniform amplitude regardless of phase or position. However, the noise entering from the port where the mirror is applied exhibits characteristics distinct from $b_F$. Due to the boundary condition imposed by the mirror, which requires the electric field to vanish at the mirror's surface, the field must be zero regardless of its amplitude. As a consequence, vacuum fluctuations must also become zero near the mirror. This boundary condition hence dictates the mode structure of all electromagnetic waves near the mirror, including vacuum modes \cite{Casimir1948, Milton2001, youn1995, wadood2019, younPRR}.
	
	Although in a standing wave the electric- and magnetic-field fluctuations show complementary spatial dependences, most practical light--matter interactions used in optical and RF detection couple predominantly to the electric field. Therefore, throughout this work we focus on the spatial modulation of the electric-field variance as the experimentally relevant noise observable, while noting that the total electromagnetic energy density is not necessarily position dependent.
	
	The spatial dependence of the noise $\langle \Delta E_1^2 \rangle$ is particularly interesting. The $(\Delta E)^2$ term's contribution to noise is spatially uniform in magnitude, although its specific instantaneous value will change with $z_1$ due to the propagation phase $k(z_1)$.
	
	The $a_{1n}$ term's contribution is modulated by $\sin(k z_1))$. This means that the magnitude of the noise contribution from the standing wave (e.g., its root mean square value over time) will vary sinusoidally with $z_1$. This implies that at positions where $k(z_1) = n\pi$ (nodes of the standing wave), the noise contribution from $a_{1n}$ will be minimal, and at $k(z_1) = (n+1/2)\pi$ (antinodes), it will be maximal.
	
	Therefore, measuring the noise power (variance) of $E_1$ as a function of $z_1$ should reveal a spatial modulation dictated by $\sin^2(k z_1)$, superimposed on a baseline noise floor from $a_{1n}$. This spatial variation in noise characteristics provides a direct experimental signature of the boundary-modified vacuum fluctuations.
	
	To understand the quantum mechanical origin of these spatially-dependent noise properties, we now turn to a more rigorous description of the electromagnetic fields in terms of quantum operators.

	\section{Quantum Mechanical Description of Fields}
	
	\subsection{Field Operators}
	
	For the traveling wave $E_b$, associated with mode $b$, the electric field operator can be expressed as \cite{Loudon2000}:
	\begin{equation}
		\hat{E}_0(t, z_1) = i \mathcal{E} \left( \hat{b} e^{-i(\omega t - k z)} - \hat{b}^\dagger e^{i(\omega t - k z)} \right). \label{eq:Eb_qm}
	\end{equation}
	Here, $E_0$ is related to a coherent state $|\alpha\rangle$, satisfying $\hat{b} |\alpha\rangle = \alpha |\alpha\rangle$, where $\alpha$ is a complex amplitude.
	
	As illustrated in Fig.~\ref{bsSetCLA}, when a beam splitter (BS) is configured with a mirror on one side, the electric field at the input port on the mirror side is reflected back to the BS after interacting with the outputs $a_1$ and $a_2$. This electric field is partially reflected and partially transmitted at the BS, resulting in the formation of a standing mode at the output by interfering with the input wave. In this case, the electric field operators are given by \cite{youn1995}:
	\begin{align}
		\hat{E}_1 &= -\frac{i}{\sqrt{2}} \mathcal{E} \Big\{ \sqrt{T} \hat{b}^\dagger e^{i (\omega t - k Z_1)} \nonumber \\
		&\quad + \hat{a}_1^\dagger e^{i (\omega t + k z_1)} - R \hat{a}_1^\dagger e^{i (\omega t - k z_1)} \nonumber \\
		&\quad - \sqrt{R T} \hat{a}_2^\dagger e^{i (\omega t - k z_1)} \Big\} + \text{H.c.}, \\
		\hat{E}_2 &= \frac{i}{\sqrt{2}} \mathcal{E} \Big\{ -\sqrt{R} \hat{b}^\dagger e^{i (\omega t - k Z_2)} \nonumber \\
		&\quad + \hat{a}_2^\dagger e^{i (\omega t + k z_2)} - T \hat{a}_2^\dagger e^{i (\omega t - k z_2)} \nonumber \\
		&\quad - \sqrt{R T} \hat{a}_1^\dagger e^{i (\omega t - k z_2)} \Big\} + \text{H.c.},
	\end{align}
	where $Z_i$ represents the propagation distance of the coherent field, and $z_i$ denotes the distance from the mirror. Additionally, if the mode entering the BS from the mirror is denoted as $c$, the modes $b$ and $c$ are related to the output modes $a_1$ and $a_2$ after passing through the BS as follows:
	\begin{align}
		\hat{a}_1 &= \sqrt{T} \hat{b} + \sqrt{R} \hat{c}, \\
		\hat{a}_2 &= -\sqrt{R} \hat{b} + \sqrt{T} \hat{c}. \label{BSeq}
	\end{align}
	
	Using these operators, we calculate the expectation values of the electric fields when mode $b$ is in a coherent state $|\alpha\rangle$, and modes $a_1$ and $a_2$ are in the vacuum state. The results are:
	\begin{align}
		\langle \hat{E}_1 \rangle &= \frac{\sqrt{T}}{\sqrt{2}} \mathcal{E} |\alpha| \sin(\omega t - k Z_1 - \theta), \\
		\langle \hat{E}_2 \rangle &= -\frac{\sqrt{R}}{\sqrt{2}} \mathcal{E} |\alpha| \sin(\omega t - k Z_2 - \theta),
	\end{align}
	where $\theta$ is the phase of the complex amplitude $\alpha$.
	
	To evaluate the fluctuations of the electric field, we compute $\langle \hat{E}_1^2 \rangle$, which yields:
	\begin{align}
		\langle \hat{E}_1^2 \rangle &= \frac{\mathcal{E}^2}{2} \Big( |\alpha|^2 T + (1 + R^2) \langle \hat{a}_1 \hat{a}_1^\dagger \rangle \nonumber \\
		&\quad + R T \langle \hat{a}_2 \hat{a}_2^\dagger \rangle + T \langle \hat{b} \hat{b}^\dagger \rangle \nonumber \\
		&\quad - 2 R \langle \hat{a}_1 \hat{a}_1^\dagger \rangle \cos(2 k z_1) \nonumber \\
		&\quad - 2 |\alpha|^2 T \cos[2 (\omega t - k Z_1 - \theta)] \Big).
	\end{align}
	At this stage, we have not yet applied the commutator relations, leaving the operators in their current form to identify the contributions of vacuum fluctuations from each mode. Using these results, the variance of the electric field fluctuations is given by:
	\begin{align}
		\langle \Delta \hat{E}_1^2 \rangle &= \frac{\mathcal{E}^2}{2} \Big( -|\alpha|^2 T + R T \langle \hat{a}_2 \hat{a}_2^\dagger \rangle \nonumber \\
		&\quad + (1 + R^2) \langle \hat{a}_1 \hat{a}_1^\dagger \rangle + T \langle \hat{b} \hat{b}^\dagger \rangle \nonumber \\
		&\quad - 2 R \langle \hat{a}_1 \hat{a}_1^\dagger \rangle \cos(2 k z_1) \Big), \\
		\langle \Delta \hat{E}_2^2 \rangle &= \frac{\mathcal{E}^2}{2} \Big( -|\alpha|^2 R + R T \langle \hat{a}_1 \hat{a}_1^\dagger \rangle \nonumber \\
		&\quad + (1 + T^2) \langle \hat{a}_2 \hat{a}_2^\dagger \rangle + R \langle \hat{b} \hat{b}^\dagger \rangle \nonumber \\
		&\quad - 2 T \langle \hat{a}_2 \hat{a}_2^\dagger \rangle \cos(2 k z_2) \Big).
	\end{align}
	
	To compute the fluctuations, we apply the vacuum fluctuation relations $\langle \hat{a}_i \hat{a}_i^\dagger \rangle =v_i ^2$ and $\langle \hat{b} \hat{b}^\dagger \rangle =|\alpha |^2+ v_b ^2$, and the condition $R + T = 1$. This yields:
	\begin{align}
		\langle \Delta \hat{E}_1^2 \rangle 
		&=\frac{ \mathcal{E}^2}{2}\Big( (1+R^2) v_1 ^2 + T (R v_2 ^2 + v_b ^2 ) \nonumber \\
		&\quad -2  R  v_1 ^2 \cos(2 k z_1) \Big), \\
		\langle \Delta \hat{E}_2^2 \rangle 
		&= \frac{\mathcal{E}^2}{2}\Big( (1+T^2) v_2 ^2 + R (T v_1 ^2 + v_b ^2 ) \nonumber \\
		&\quad -2  T  v_2 ^2 \cos(2 k z_2) \Big).
	\end{align}
	
	\begin{figure}[htb]
		\centering
		\includegraphics[width=0.6\linewidth]{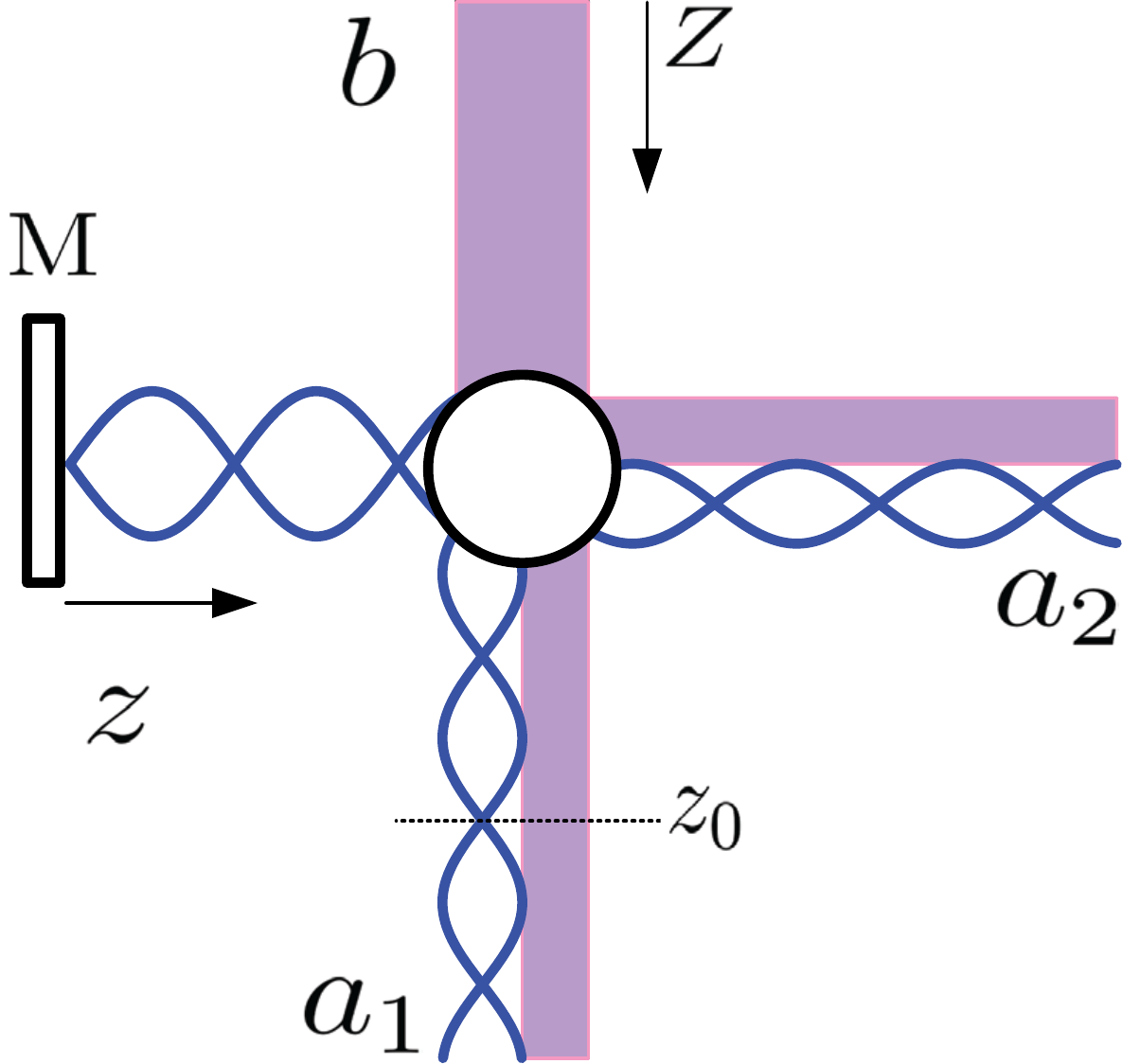}
		\caption{Quantum noise characteristics divided by a beam splitter (BS). M denotes a mirror, and the two output ports of the BS are labeled as $a_1$ and $a_2$. At the mirror side, the noise forms a standing wave due to the boundary condition. The figure shows the case where $T = \frac{1}{2}$.}
		\label{bs_setup}
	\end{figure}
	
	Following the approach in our semiclassical model, where we normalized the vacuum fluctuations carried by the coherent state as a sum over all phases of forward- and backward-propagating modes, we can similarly analyze the quantum mechanical field noise of the coherent state entering the beam splitter. Using Eq.~(\ref{delCo}) and expressing the total noise as $v_b^2 + v_1^2 T + v_2^2 R$, we obtain:
	\begin{align}
		\langle \Delta \hat{E}_1^2 \rangle &= \mathcal{E}^2\Big( \frac{T}{2} (v_b^2 + R v_2^2 + T v_1^2) \nonumber \\
		&\quad + 2 R v_1^2 \sin^2(k z_1) \Big), \label{DelE1} \\
		\langle \Delta \hat{E}_2^2 \rangle &= \mathcal{E}^2\Big( \frac{R}{2} (v_b^2 + R v_2^2 + T v_1^2) \nonumber \\
		&\quad + 2 T v_2^2 \sin^2(k z_2) \Big).
	\end{align}
	
	These expressions demonstrate, consistent with our semiclassical approach, that the field fluctuations can be decomposed as:
	\begin{align}
		\Delta E_1^2 &= T (\text{QNL from } E_0) \nonumber \\
		&\quad + (\text{QNL from } a_2) \left(2 R \sin^2(k z_1) \right),  \\
		\Delta E_2^2 &= R (\text{QNL from } E_0) \nonumber \\
		&\quad + (\text{QNL from } a_1) \left(2T \sin^2(k z_2) \right), \label{delE1}
	\end{align}
	where QNL denotes the quantum noise level contribution from the respective sources.
	
	This final form clearly shows which mode contributes to the vacuum noise. 
	This confirms the semi-classical prediction: the noise power measured in $E_1$ will vary spatially, with minima occurring at the nodes of the standing wave ($k z_1 = n\pi$) where the contribution from $a_1$'s vacuum fluctuations is suppressed by its mode function, and maxima at the antinodes ($k z_1 = (n+1/2)\pi$).

	\section{Measurement and Feedback Control of Vacuum Fluctuations}
	
	In the previous section, we analyzed the electric field fluctuations at the output ports of a beam splitter (BS) with a mirror at one input port and coherent light incident on the other. Here, we explore a method to stabilize the output of electromagnetic wave generators by extracting a portion of the output via a BS, measuring its fluctuations, and employing this information in a feedback loop. While classical noise can be effectively suppressed within the feedback loop bandwidth using such techniques, quantum mechanical noise, due to its fundamental nature, cannot be eliminated in this manner.
	
	The photocurrent operator is defined as $\hat{I} = \hat{E}^{(-)} \hat{E}^{(+)}$ \cite{Loudon2000}, where
	\begin{equation}
		\hat{E}^{(-)} = \mathcal{E} \left( \hat{a}_{1F}^{\dagger} e^{i(\omega t - k z_1)} + \hat{a}_{1B}^{\dagger} e^{i(\omega t + k z_1)} \right).
	\end{equation}
	Here, $\hat{a}_{1F}$ and $\hat{a}_{1B}$ represent the field operators for modes propagating in the $+z$ and $-z$ directions, respectively, facilitating comparison with the mirror-coupled BS configuration.
	
	Consider a BS with input ports labeled as modes $b$ and $c$, connected to output ports $a_1$ and $a_2$ via the BS relations in Eq.~(\ref{BSeq}). We examine the scenario where mode $b_F$ is in a coherent state $|\alpha\rangle_{b_F}$, while all other modes are in vacuum states. The fluctuations in the bidirectional modes at ports $b$ and $c$ are given by:
	\begin{align}
		\langle \hat{b}_F \hat{b}_F^{\dagger} \rangle &= |\alpha|^2 + v_{bF}^2, \quad
		\langle \hat{b}_B \hat{b}_B^{\dagger} \rangle = v_{bB}^2, \\
		\langle \hat{c}_F \hat{c}_F^{\dagger} \rangle &= v_{cF}^2, \quad
		\langle \hat{c}_B \hat{c}_B^{\dagger} \rangle = v_{cB}^2,
	\end{align}
	where $v_{ij}^2$ denotes the vacuum fluctuations in the respective modes.
	
	The variance of the photocurrent measured at port $a_1$ is:
	\begin{align}
		\langle (\Delta \hat{I}_1)^2 \rangle &= |\alpha|^2 T \Big[ R(v_{cB}^2 + v_{cF}^2) \nonumber \\
		&\quad + T(v_{bB}^2 + v_{bF}^2) \Big].
	\end{align}
	This expression reveals that when the BS extracts a fraction $T |\alpha|^2$ of the input intensity for measurement, vacuum fluctuations from the unused input port (mode $c$) are inevitably included, scaled by the reflectance $R$. Consequently, even ideal feedback control based on measurements at mode $a_1$ cannot reduce the fluctuations in mode $a_2$ below the quantum noise limit \cite{Zhang2017, Yamamoto1986,Jacobs}. 
	
	However, placing a mirror at the vacuum input port, as depicted in Fig.~\ref{bs_setup}, modifies the photocurrent variance at mode $a_1$, yielding, analogous to Eq.~(\ref{DelE1}):
	\begin{align}
		\langle (\Delta \hat{I}_1)^2 \rangle &=  
		T |\alpha|^2 \Big[ T \left( v_b^2 + R v_2^2 + T v_1^2 \right) \nonumber \\
		&\quad + 2 R v_1^2 \sin^2(k z_1) \Big].
	\end{align}
	Here, $v_b^2$ represents vacuum fluctuations propagating in the $+z$ direction in mode $b$, while $v_1^2$ and $v_2^2$ correspond to vacuum fluctuations propagating from the output ports back toward the input, related to the backward-propagating vacuum fluctuations opposite to the coherent light in mode $b$. The term $(v_b^2 + R v_2^2 + T v_1^2)$ thus encapsulates the fluctuations associated with the electric field in mode $b$.
	
	The mirror at the unused input port alters the vacuum fluctuations entering the system, introducing a position-dependent term, $\sin^2(k z_1)$, which arises from interference between the reflected vacuum modes and the input coherent state. By tuning the mirror position (i.e., adjusting $z_1$), the contribution of certain vacuum fluctuations can be minimized, optimizing feedback stabilization. Nevertheless, the fundamental quantum noise limit remains insurmountable due to the intrinsic quantum nature of the electromagnetic field.
	
	This mirror configuration establishes a standing wave pattern, creating spatial nodes where the vacuum noise contribution from the mirrored input port vanishes. At specific positions (e.g., $z_1 = z_0$), the standing wave exhibits a node, eliminating vacuum noise associated with mode $E_1$. 
	At these nodal points, measurements at $E_1$ predominantly sample the vacuum fluctuations carried by the coherent input mode ($b$), because the mirror-side standing-wave contribution vanishes at nodes. However, standing wave vacuum fluctuations from $E_2$ persist, manifesting as a spatial modulation of the vacuum noise with an intensity profile proportional to $\sin^2(k z_2)$, where $z_2$ is the optical path length from the BS to the mirror.
	
	In the language of quantum feedback, noise reduction produced by negative feedback is often ``in-loop'' (sometimes called squashing): the field used for the error signal can exhibit reduced fluctuations, while an independent out-of-loop output does not necessarily show the same reduction because additional, uncorrelated vacuum fields enter at other ports \cite{Yamamoto1986,Youn1994,Zhang2017}. The role of the mirror in the present scheme is precisely to engineer the vacuum mode structure so that, at standing-wave nodes, the mirror-side contribution is (ideally) suppressed; under this condition the feedback can act on the dominant remaining noise source, allowing sub-SQL noise to appear also at the other (out-of-loop) port \emph{at corresponding node positions}, subject to the practical limitations discussed below.
	
	Implementing this scheme experimentally imposes stringent requirements. First, the electromagnetic radiation must be highly monochromatic to sustain coherent standing wave patterns. Second, precise mode matching between incident and reflected fields is essential for effective interference. Third, the detector must achieve spatial resolution along the propagation direction comparable to or better than the radiation wavelength \cite{YounMode}. This last requirement is particularly challenging in the visible light regime, where sub-wavelength resolution along the propagation axis is technically demanding, rendering complete elimination of vacuum fluctuations from port $E_v$ impractical. However, the underlying principles are universal across the electromagnetic spectrum, suggesting greater feasibility at other frequencies.
	
	Quantitative analyses of how finite source bandwidth and finite spatial averaging degrade the node-like suppression were presented in Ref.~\cite{youn1995} (vacuum-field modulation in the BS--mirror system) and in Ref.~\cite{YounMode} (sub-Poisson/feedback performance including realistic detection considerations). These works show that the observable reduction can be substantially smeared when the detector averages the field over an axial extent that is not small compared to the wavelength, or when the coherence length is not large compared to the relevant path differences; consequently, in the visible regime the requirements become extremely demanding, whereas in longer-wavelength regimes they become comparatively accessible.
	
	Importantly, Ref.~\cite{younPRR} reported a room-temperature MHz-domain experiment in which placing a reflecting termination at the BS port suppresses the measured \emph{thermal} noise at specific node positions. While this does not yet constitute a direct vacuum-noise measurement, thermal noise in such guided-wave implementations is generated in the same set of allowed electromagnetic modes; thus the observed position-dependent suppression provides indirect, but practical, evidence that cooling the system would allow the corresponding vacuum fluctuations to be reduced at specific locations as predicted here.
	
	Recent experiments in the radio frequency domain have demonstrated the viability of this approach. Using BNC cables to guide MHz electromagnetic waves, the spatial distribution of vacuum fluctuations near a reflecting termination has been indirectly verified \cite{younPRR}, providing a proof-of-principle for implementations where the required spatial resolution is more achievable.
	
	The ability of feedback to control vacuum-level fluctuations may seem counterintuitive, but it is rooted in the fact that Poissonian statistics of light intensity originate from vacuum fluctuations. Thus, reducing intensity fluctuations inherently involves manipulating vacuum noise. In high-intensity optical fields, feedback-based intensity stabilization effectively controls vacuum fluctuations. Previous experiments with semiconductor lasers explored multi-mode vacuum feedback schemes but, lacking the standing wave configuration described here, could not achieve sub-shot-noise operation due to their inability to distinguish between different vacuum noise contributions \cite{Youn1994}.
	
	Technical challenges in this feedback scheme include managing time delays in the feedback loop, which may be mitigated using feedforward techniques or operating in regimes where the feedback bandwidth significantly exceeds noise frequencies. Additionally, maintaining the stability of the standing wave pattern requires precise control of optical path lengths and isolation from environmental perturbations.
	
	The implications of this work extend beyond noise reduction in optical systems. The ability to selectively measure and manipulate vacuum fluctuations from specific spatial modes offers new avenues for quantum state engineering, with potential applications in quantum information processing and precision metrology.

	\section{Conclusion}
	
	We have presented a comprehensive analysis of the electric field $\hat{E}_1$ resulting from the mixing of a coherent traveling wave and a vacuum standing wave at a beam splitter. Both semi-classical and quantum mechanical treatments consistently predict that the noise characteristics of $\hat{E}_1$ will exhibit a distinctive spatial modulation pattern dictated by the standing wave's mode function, superimposed upon the uniform vacuum noise contribution from the traveling wave. 
	
	Remarkably, by simply placing a mirror at one output port of the beam splitter, we demonstrate that the noise at specific positions such as $z_1 = z_0$ can be reduced below the standard quantum limit---the fundamental noise floor typically associated with coherent states. This sub-shot-noise behavior occurs at the nodes of the vacuum standing wave, which is formed by placing a mirror at one of the input ports of the beam splitter.
	
	Furthermore, by implementing an active feedback loop based on the noise measurements at $\hat{E}_1$ to control the input laser field $\hat{E}_b$, amplitude squeezing can be achieved in the input field. This squeezing effect propagates through the beam splitter to the output port $\hat{E}_2$, where sub-quantum-noise-limit operation can be observed at positions satisfying $z_2 = n\lambda/2$. The resulting squeezed field $\hat{E}_2$ exhibits reduced amplitude fluctuations while maintaining a spatial noise profile that reflects the underlying standing wave structure, with maximum noise suppression occurring precisely at the standing wave nodes.
	
	To avoid confusion, we emphasize that feedback-based noise reduction is, in general, an in-loop effect (squashing) and does not automatically imply freely propagating out-of-loop squeezing at all positions. In the present scheme, out-of-loop sub-SQL behavior is enabled only in a mode-selective manner, i.e., at locations where the mirror-engineered standing-wave contribution is suppressed and the feedback can effectively address the remaining dominant noise source.
	
	The central insight of this work represents a conceptual breakthrough in quantum noise control: quantum noise can be suppressed by engineering the spatial ``mode'' of light, rather than by manipulating its quantum ``wave function'' through nonlinear optical processes. This distinction---between the mode structure dictated by boundary conditions and the quantum state described by the wave function---has often been conflated in the literature, yet it is precisely this distinction that enables our remarkably simple approach. By merely introducing a mirror to impose appropriate boundary conditions, we reshape the vacuum mode structure to create nodes of exceptionally low fluctuation, achieving sub-shot-noise performance without any nonlinear operations. This method stands in stark contrast to conventional squeezed-state generation techniques, which require sophisticated nonlinear optical elements such as parametric amplifiers or four-wave mixing media. The simplicity and elegance of our approach---requiring only a single mirror and a beam splitter---makes it highly practical and broadly accessible for a wide range of precision measurement applications.
	
	This work illuminates the spatial mode structure and quantum noise in optical systems, revealing novel pathways for precise control and manipulation of electromagnetic fields at the quantum limit. The ability to selectively isolate and manipulate vacuum fluctuations from specific spatial modes through this simple mirror-and-beam-splitter configuration represents a significant advancement in quantum optics.
	
	Most importantly, the theoretical framework and experimental principles presented here are universally applicable across the entire electromagnetic spectrum---from radio waves to gamma rays. This broad applicability ensures that our findings will provide pivotal insights for all research domains where electromagnetic noise plays a critical role, including gravitational wave detection, quantum metrology, precision spectroscopy, quantum information processing, and the development of ultra-stable frequency standards. As detection technologies continue to advance across different frequency regimes, we anticipate that these principles will enable new generations of quantum-limited measurements and quantum state engineering protocols.

	\begin{acknowledgments}
		The publication of this work was supported by Creation of the quantum information science R\&D ecosystem (based on human resources, RS-2023-00256050) through the National Research Foundation of Korea (NRF) funded by the Korean government (Ministry of Science and ICT).
	\end{acknowledgments}

\end{document}